\documentclass[aps,prb,twocolumn,showpacs,floatfix]{revtex4}
\usepackage{graphicx}
\usepackage{dcolumn}\usepackage{amsmath}

\newcommand{\beq}{\begin{equation}}
\newcommand{\eeq}{\end{equation}}
\newcommand{\bea}{\begin{eqnarray}}
\newcommand{\eea}{\end{eqnarray}}
\newcommand{\be}{\begin{equation}}
\newcommand{\ee}{\end{equation}}
\newcommand{\pdag}{{\phantom{\dagger}}}
\newcommand{\im}{{\rm Im\;}}

\begin{document}
\draft

\title{Two-stage Kondo effect  
in side-coupled quantum dots: Renormalized perturbative scaling theory and 
Numerical Renormalization Group analysis}
\author{Chung-Hou Chung$^1,^3$, Gergely Zarand$^1,^2$, and Peter W\"olfle$^1,^4$}
\affiliation{
$^1$ Institut f\"ur Theorie der Kondensierten
Materie, Universit\"at Karlsruhe, 76128 Karlsruhe, Germany\\
$^2$ Department of Theoretical Physics,
Institute of Physics,
Budapest University of Technology and Economics, Budapest, Hungary\\
$^3$ Electrophysics Department, National Chiao-Tung University, HsinChu, Taiwan, R.O.C.\\
$^4$ Institut f\"ur Nanotechnologie, Forschungszentrum Karlsruhe, 76026 Karlsruhe, Germany}  
\date{\today}
\begin{abstract}
 We study numerically and analytically the dynamical (AC) conductance through
 a two-dot system,  where only one of the dots is  coupled to the leads but it
 is also side-coupled 
to the other dot  through an antiferromagnetic exchange (RKKY) interaction. In this case
the RKKY interaction  gives rise to a ``two-stage Kondo
effect''  where the two spins are screened by two consecutive Kondo effects. We formulate a renormalized 
scaling theory that captures remarkably well the cross-over from 
the strongly conductive correlated regime to the low temperature low conductance state.
Our analytical formulas agree well  with our numerical renormalization group results. 
The frequency-dependent current noise spectrum is also discussed. 
\end{abstract}
\maketitle

{\em Introduction.}
The Kondo effect\cite{hewson} in
semiconductor quantum dots has attracted significant theoretical and
experimental interest  in recent years.\cite{Kondo-popular,kondo,kondo-theo} 
In a  dot that is only weakly coupled to leads, charge fluctuations are typically 
suppressed due to Coulomb blockade \cite{Coulomb-blockade}. 
However, if the dot has an odd number of electrons then 
the spin of this electron interacts antiferromagnetically 
with the spin of the conduction electrons in the leads,  
and at low temperatures it is screened through a Kondo effect.   
The formation of this Kondo state typically leads to an enhancement of 
the conductance at low bias voltages. 

Recently, in the quest for designing multiple quantum dot systems with
tunable spin control, which can be used in  spintronics and quantum information
processing, double quantum dots have become the focus of interest.\cite{Craig,heersche}
In these systems, when both dots are tuned to the single spin 
regime, an effective spin-spin interaction known as the
Ruderman-Kittel-Kasuya-Yoshida (RKKY)\cite{RKKY} interaction is mediated 
between the two dots by the  conduction electrons.\cite{Craig} This RKKY coupling
competes with the Kondo effect in these systems.  In the case where the two
dots are coupled to two separate electrodes, an antiferromagnetic 
RKKY interaction  leads to a cross-over  between the Kondo and  RKKY
regimes\cite{Craig,simon,vavilov,chung1,chung2}: a sufficiently 
strong  antiferromagnetic RKKY coupling will lock the spins of the 
two dots into a singlet, and thereby suppress the Kondo effect,  
while the Kondo effect persists for weak RKKY interactions. 
This picture is slightly modified under non-equilibrium conditions: Then,
even for 
strong RKKY interaction, the Kondo effect is partially restored by the finite bias voltage 
allowing for triplet excitations.\cite{koerting}
Part of the rich physics has been studied previously in the framework of  two impurity
Kondo\cite{2impkondo} and Anderson impurity\cite{2impAnderson} models, and the 
singlet-triplet cross-over has also been recently   studied  experimentally 
by Craig {\it et   al.},\cite{Craig} who observed a the Kondo resonance for weak RKKY couplings 
and a splitting of the Kondo resonance for large RKKY interactions.
Significant theoretical and experimental effort has been devoted
to the Kondo-RKKY transition in these double quantum dot systems,
and related multi-orbital systems with two Kondo-screening channels have
also been studied extensively in recent years by different theoretical 
approaches.\cite{kroha}

\begin{figure}[bp]
\begin{center}
\includegraphics[width=0.8\columnwidth,clip]{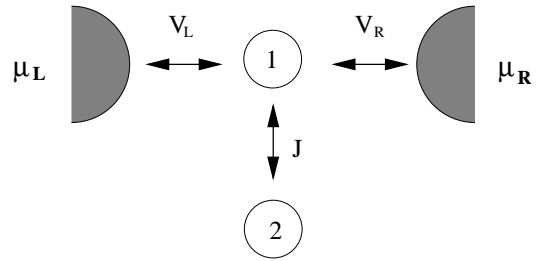}
\end{center}
\vspace{-0.1cm}
\caption{Sketch of the two side-coupled quantum dots.  
Electrons tunnel from only one of them to the leads and they interact through 
an effective exchange interaction, $J$.}
\label{set}
\end{figure}

Rather than studying the usual singlet-triplet transition itself, 
in the present paper we shall focus on an even simpler but equally
interesting arrangement, where only one of the dots is coupled to external
leads, 
but the two dots are still side-coupled to each other (see Fig.~\ref{set}). 
Unlike the usual four lead set-up, where the cross-over between an RKKY and
Kondo regimes is governed by a quantum critical point corresponding to a finite value of the 
RKKY coupling $J$,\cite{chung1,chung2}  in this side-coupled
 system the side-coupled spin is always screened for 
any positive $J$, and  a Kosterlitz-Thouless-type quantum phase transition 
and a two-stage Kondo effect occurs.\cite{grempel,vojta} 
We remark that 
this transition is essentially the two-dot 
analogue  of the singlet-triplet quantum phase 
transition found in single dot devices, when the two dot levels
are coupled to a single conduction electron mode in the 
leads.\cite{Scholler,Kogan}

Although many interesting results have been obtained recently for this 
side-coupled system,\cite{grempel,vojta}  there is still a lack of a 
detailed analytical and numerical understanding   
of the two-stage Kondo screening processes in the vicinity of the quantum phase 
transition. The main goal of the present work is to have some more detailed 
theoretical control and understanding of the transport properties of this
transition. We shall reach this goal by combining numerical renormalization
 group\cite{nrg} and renormalized perturbative scaling approaches. As we
 shall see, the latter relatively simple analytical framework is able to account for the 
numerical results over a wide range of energy scales, and together with Fermi
liquid theory, it provides a reliable theoretical framework to understand the
low energy cross-over. 

We mostly focus on the  T-matrix, but also present results on 
the linear AC conductance and the equilibrium current fluctuations. Both of
these  quantities can be observed experimentally and show features 
characteristic of the two-stage Kondo effect.

{\em The Model.} For the numerical calculations we shall describe the 
two side-coupled dots in Fig.~\ref{set}  by an 
Anderson-like model. The two isolated dots are described by the Hamiltonian 
\bea 
H_{DD} &=& 
\sum_{i=1,2} \frac{U_i}{2} (N_i- n_{gi})^2\; + J \, {\bf S}_1\, {\bf S}_2,
\label{eq:HDD}
\eea
where  $i=1,2$ labels the two
dots, $N_{i}=\sum_{\sigma }d_{i\sigma }^{\dagger }d_{i\sigma }^{\pdag}$ is
the number of electrons occupying dot $i$ and 
${\bf S}_{i}=(1/2)\sum_{\sigma \sigma ^{\prime}}d_{i\sigma }^{\dagger }{\bf \sigma}_{\sigma \sigma
^{\prime }}d_{i\sigma ^{\prime }}^{\pdag}$ is their spin. 
Each dot is subject to a {\em charging energy}, $U_1\approx U_2 = U = E_C$,  and the two dots 
are coupled by an exchange coupling, which is assumed to be
antiferromagnetic, $J>0$. The $n_{gi}$ in Eq.~\eqref{eq:HDD} denote
dimensionless gate voltages that set the occupation numbers, $\langle
N_i\rangle$. 
In the rest of the paper, we restrict ourself to the case of particle-hole
symmetry, $n_{g1}=n_{g2}=1$. However, this assumption is expected to be
unimportant as long as $n_{g1}\approx n_{g2}\approx 1$. 

The coupling of dot 1 to the leads is modeled by the usual tunneling Hamiltonian, 
\bea
H_t &=& \sum_{\alpha=L,R}\sum_{\epsilon,\sigma } ( V_{\alpha} c^{\dagger}_{\alpha \epsilon \sigma} d_{1,\sigma} + h.c.)\;.
\eea
Here, $V_L$ and $V_R$   denote the tunneling amplitudes to  the left and right
leads, respectively, and  $c^{\dagger}_{\alpha \epsilon \sigma}$ creates an electron 
in lead $\alpha=L,R$ with spin $\sigma$ and energy $\epsilon$. 
This tunnel coupling leads to a broadening of the level on dot 1, the width of
which is given  by  $\Gamma =\Gamma_{L}+\Gamma _{R}= 2\pi (V_L^{2}\varrho_{L} + V_R^{2}\varrho_{R})$, with  
$\varrho_{L/R}$ the density  of states in the leads.

{\em AC conductance and noise:} 
Our main goal is to determine AC transport properties in the 
linear response regime and see how the two-stage effect appears in these quantities. 
Fortunately, since there is no charge transfer between the two dots, 
the derivation of Ref.~\onlinecite{sindel} carries over to our case, and the  
real part of the optical conductance is simply given by 
\begin{equation}
G^{\prime }(\omega )=  \frac {G_0}{4\;\omega}\sum_\sigma \int d\omega' [ f(\omega' - \omega)
- f(\omega' + \omega )] \;T_\sigma (\omega'),
\label{eq:ReG}
\end{equation}
where $T(\omega )$ is ``the transmission probability'' at energy $\omega$, and 
\be
G_0 = \frac{2e^2}{h}\frac{4\;\Gamma_{L}\;\Gamma _{R} }{(\Gamma_{L}+\Gamma _{R})^2}
\ee 
denotes the maximum conductance through the dot. The
transmission coefficient  $T_\sigma(\omega)$ appearing in Eq.~\eqref{eq:ReG}
can be expressed as 
\begin{equation}
T_\sigma(\omega )= - \Gamma \; {\rm Im\;} G_{11\sigma }(\omega ),  
\label{transmission}
\end{equation}
where we have introduced the retarded  Green's function on dot $1$: 
$G_{11\sigma }(t)=-i\theta (t)\langle \{d_{1\sigma }^{\pdag}(t),d_{1\sigma }^{\dagger }\}\rangle $. 
This Green function can be computed accurately using  numerical
renormalization group methods,\cite{nrg} but as we shall see, substantial
analytical progress can be made by a special version of renormalized perturbation theory. 
Although Eq.~\eqref{transmission} is valid at any temperature, in the following 
we shall focus our attention to the high frequency regime, $\omega \gg  T$,
and set the temperature to zero, $T=0$.  

The  noise spectrum of the device is also of experimental relevance. This is  defined as
\begin{equation}
C(\omega )=\int_{-\infty }^{\infty }dte^{{\it i}\omega t}[\langle
I(0)I(t)\rangle -\langle I\rangle ^{2}].
\end{equation}
At equilibrium, this is simply related to the linear 
conductance $G^{\prime }(\omega )$ by the fluctuation-dissipation theorem,\cite{sindel} 
\begin{equation}
C(\omega )=\frac{2\hbar \omega }{exp(\hbar \omega /kT)-1}\;G^{\prime }(\omega )\;.
\end{equation}
This formula simplifies further at $T=0$ temperature 
to  $C(\omega )=2 \hbar |\omega| \;G^{\prime }(\omega )\;\theta(-\omega)$. 
Clearly, to determine the equilibrium noise spectrum of the device and its 
AC conductance only the Green's function of the first dot needs to be determined.

{\em Renormalized perturbative scaling:} Before presenting our numerical
results, let us reach some analytical understanding of the physics of the
side-coupled dot in the limit  $J\to 0$. In this regime, ``two-stage Kondo
screening'' takes place\cite{grempel}: In the first stage, the
spin of the dot $1$ gets screened below the  Kondo temperature $T_{K}\approx
D\;e^{-\pi U/\Gamma }$, where the high energy cut-off $D$ denotes the effective
half bandwidth of the conduction electrons.\cite{footnote}
 Clearly, for the Kondo effect 
to take place $J \ll T_{K}$ is required, otherwise the two spins are locked
together to a singlet before the Kondo effect can take place.
 Then below the Kondo scale $T_K$ the electron on the first dot is
dissolved in the conduction electron sea of the leads, and presents an
effective  fermionic bath for the 
electron on the second dot. Since the coupling between the two dots is 
antiferromagnetic, another Kondo effect shall take place  at a much 
smaller energy scale, $T^\star$, where the spin of the 
second dot is also screened. Our aim is to understand the formation 
of this second Kondo singlet in detail. 

We first observe that in the regime of interest, $\omega \ll T_{K}$ 
the Matsubara Green's function of dot 1 can be approximated by the resonant 
level expression,\cite{hewson}
\begin{equation}
G_{d1}^{0}({\it i}\omega _{n})=\frac{z}{{\it i}\omega _{n}+{\it i}\;\tilde T_K
sgn(\omega _{n})}\;,
  \label{Gd10}
\end{equation}%
where $z= c\; \frac{T_{K}}{\Gamma }$ denotes the quasiparticle weight at the
Fermi energy, and $\tilde T_K = z\Gamma = c\; T_K$ is an energy of the 
order of the Kondo temperature, $T_K$. 
The precise value of the universal constant $c$ relating $T_K$ and $\tilde
T_K$  depends on the
definition of $T_K$. Throughout this paper we shall define $T_K$ as the
half-width of the transmission $T(\omega)$. Then from fitting the NRG data we
get $c\approx 0.5$.  Note that the Lorentzian representing the Kondo resonance has a
very small spectral weight, $z\ll 1$, and most of the spectral weight goes
to the Hubbard peaks.  

Fermi liquid theory and the basic principles of renormalization group 
also imply that the exchange interaction between the 
dot spins is renormalized by  the same $z$-factor, i.e., in the regime 
$\omega\ll T_K$ the effective RKKY interaction reads
\be 
H_{\rm RKKY} \to  \frac {\tilde J} z  \;{\bf S}_1\, {\bf S}_2\;.
\label{eq:renormRKKY}
\ee
In a first approximation one would think that, $\tilde J = J$, however, the
fact that ${\rm Im}G_{d1}$ has a large logarithmic tail above $T_K$ leads to a
slight renormalization of this relation. From a fitting of the numerical data
shown later, we obtain the approximate relation $\tilde J \approx 1.1\; J$.

\begin{figure}[tp]
\begin{center}
\includegraphics[width=0.9\columnwidth,clip]{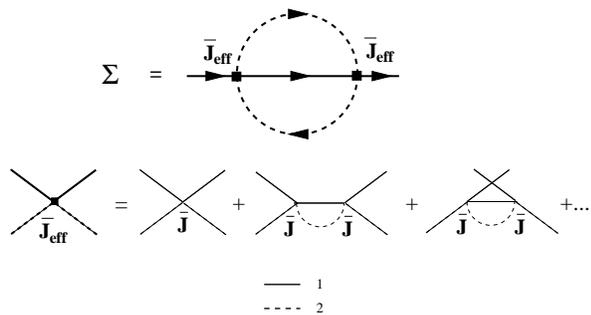}
\end{center}
\vspace{-0.1cm}
\caption{Top: leading logarithmic contribution to the self energy of 
the first dot's Green's function. Solid  lines represents the propagator $G_{d1}^{0}$
of  quantum  dot 1, given by Eq.~\eqref{Gd10},  while dashed
lines denote the pseudofermion propagator associated with the side-coupled 
dot's spin, $S_2$. Full squares stand for the full leading logarithmic 
vertex function, given in the lower part of the figure. Crossings 
of dashed and continuous lines correspond to an interaction through the 
renormalized RKKY interaction, $J_{\rm eff }=\tilde J/z$, given by 
Eq.~\eqref{eq:renormRKKY}.
 }
\label{self}
\end{figure}
We are now in the position to develop a  perturbative scaling 
theory in the small coupling, $\tilde J$. To do this, we used Abrikosov's 
pseudofermion representation to compute the second order 
self energy and vertex  corrections, shown in  in Fig.\ref{self}.
The dimensionless vertex function is given by the following expression: 
\be
\gamma(\omega) \equiv  \varrho(\omega) \Gamma(\omega)
= \hat  \varrho(\omega) \tilde J + 
(\hat  \varrho(\omega) \tilde J)^2 
\log \left( \frac{\tilde T_K }{-\omega}\right)
\dots\;,  
\label{eq:logcorr}
\ee
where $\tilde T_K = c\;T_K$ is the effective width of the Kondo resonance,
$\omega$ is the energy of the incoming electron, 
and $\hat  \varrho(\omega)$ denotes the rescaled effective density of states 
of dot 1, serving as a many-body reservoir for dot 2, 
\be 
\hat  \varrho(\omega)\equiv   
\varrho(\omega)/z = \frac {\tilde T_K} 
{\pi (\omega^2 + {\tilde T_K}^2 ) }\;.
\ee
Eq.~\eqref{eq:logcorr} is only of logarithmic accuracy, irrelevant terms of 
order $\omega/\tilde T_K$ have been neglected.

The Kondo temperature $T_K \sim \tilde T_K$ appears in Eq.~\eqref{eq:logcorr}
as a high energy cut-off. We can therefore perform a scaling transformation 
by reducing this cut-off, $\tilde T_K \to \tilde T_K'$, and  
requiring the invariance of the vertex function 
for frequencies $\omega < \tilde T_K $ at the same time. This transformation 
sums up all leading logarithmic diagrams and leads to the following scaling equation
\be 
\frac{d (\hat  \varrho(\omega) \tilde J)}{dl} = ({\hat  \varrho}(\omega) \tilde J)^2\;,
\label{eq:scaling}
\ee
with the scaling variable defined as $l\equiv \log (\tilde T_K / \tilde
T_K')$. Integrating this differential equation up to 
$l\equiv \log (\tilde T_K / \omega)$,  one obtains the dimensionless vertex 
function in the leading logarithmic approximation:  
\be
\gamma(\omega,\tilde T_K) = 
\frac{\hat  \varrho(\omega) \tilde J} {1 + \hat  \varrho(\omega) \tilde J
  \;\log(-\omega/ \tilde T_K)}\;.\ee
This equation can be rewritten with a little algebra as
\be 
\gamma(\omega,\tilde T_K) = 
\frac1 {\frac {\omega^2}{\tilde T_K^2} \log\frac {\tilde T_K}{T^*}
+ \log\frac {-\omega}{T^*}}\;,
\ee
with the second scale $T^*$ defined as 
\be 
T^* = \tilde T_K \; \exp(-\pi\;\tilde T_K/\tilde J)\;.
\label{eq:Tstar}
\ee
This scale also appears in the slave boson approach of Ref.~\onlinecite{grempel},
however, the latter approach does not account for the logarithmic 
corrections, which are of our main interest here.
Clearly, the dimensionless vertex diverges at an energy, $\omega\approx
T^\ast$, implying that the effective RKKY interaction becomes dominant 
below this scale and diverges in the limit  $\omega\to 0$.\cite{nrg} 
In other words, below $T^\ast$ a singlet is formed from the two dot spins, and 
the scale $T^\ast$ can thus be viewed as the effective singlet-triplet
binding energy.  Note that the spin-singlet is always the ground state for any arbitrary RKKY
coupling $J>0$. 

The point $\tilde J=0 $ is special: It separates the
ferromagnetic phase ($J<0$) from the antiferromagnetic phase discussed so far ($J>0$). 
While in the antiferromagnetic phase the two dot spins are locked into a
singlet, in the ferromagnetic  phase the spins of the dots are bound to a
triplet,  which is then partially screened by the lead electrons, and
correspondingly, the ground state is a doublet for $J<0$. Thus for $J<0$ the side-coupled system has a residual
entropy. At the critical point, $J=0$, the scale $T^\ast$ diverges exponentially, 
corresponding to a Kosterlitz-Thouless phase transition
between these two states.\cite{vojta,grempel}

The second order self-energy correction to the retarded Green's function $G^0_{d1}$ simply 
gives the expression 
\be
\Sigma (\omega )= S(S+1)\frac {\tilde J ^{2}}{4 z} \; \frac 1{\omega + i \tilde
T_K}\;.
\ee
where S=1/2. Note that this correction scales as $\sim J^2/(T_K\;z)$ and, although it looks to
be very large at a first sight,   it is actually small compared to
  $(G^0_{d1})^{-1}\sim  T_K/z$ as long as $J\ll T_K$.
In the leading approximation, this self-energy results in the following 
Green's function:
\be
\Gamma \;G^{(2)}_d(\omega) = \frac{\tilde T_K}
{\omega + i  \;\tilde T_K - 
 \frac{\bar J ^2 \; S(S+1)}{4}\;  \frac 1{\omega + i\tilde T_K}} \;. 
\label{eq:G^2_d}
\ee

\begin{widetext}
Summing up the leading logarithmic corrections to the self-energy simply amounts to 
replacing $\tilde J$ in Eq.~\eqref{eq:G^2_d} by
$\gamma(\omega)/\hat \varrho(\omega)$, and thereby results in the following
transmission coefficient, 
\be
T_\sigma(\omega) = 
- \im \left\{
\frac {\tilde T_K^3}
{
\omega \left( \tilde T_K^2  - \frac14 S(S+1) \pi^2 (\omega^2 + \tilde T_K^2)
  \gamma^2(\omega)\right)
+ i \tilde T_K \left( \tilde T_K^2  + \frac14 S(S+1) \pi^2 (\omega^2 + \tilde T_K^2)
  \gamma^2(\omega)\right)
}
\right \}\;.
\label{eq:G_d}
\ee
\end{widetext}
The logarithmic corrections hidden in $\gamma$ result in  the formation of a dip 
in the spectral density of dot $1$, corresponding to a 
{\em suppression} of the transmission coefficients at energies 
$\omega\sim T^\ast \ll T_K$. 
This dip in $T_\sigma(\omega)$   also implies the
appearance of a dip in the AC conductance discussed  later, and is a
clear signature of the formation of a singlet ground state. 
Physically, it is a consequence of the fact that electrons promoted from
one side of the device to the other must first break up the singlet 
of energy $T^\ast$ formed by the two dot spins. Clearly, electrons of energy 
$\omega< T^\ast$ are not energetic enough   to break up this singlet and
therefore their transport is suppressed.

The logarithmic approximation breaks down below  $\omega \sim T^\ast$. 
There a Fermi liquid is formed and $T_\sigma(\omega)$ scales as
\be
T_\sigma(\omega) \approx a  + b \frac {\omega^2}{(T^\ast)^2}
\;,
\phantom{nnn} (\omega \ll T^\ast )
\;. 
\ee
The constant $a$ vanishes for  electron-hole symmetry
and remains typically small unless electron-hole symmetry is dramatically broken, 
while the coefficient  $b$ is a number of the order of unity.

\begin{figure}[bp]
\begin{center}
\vspace{0.2cm}
\includegraphics[width=0.9\columnwidth,clip]{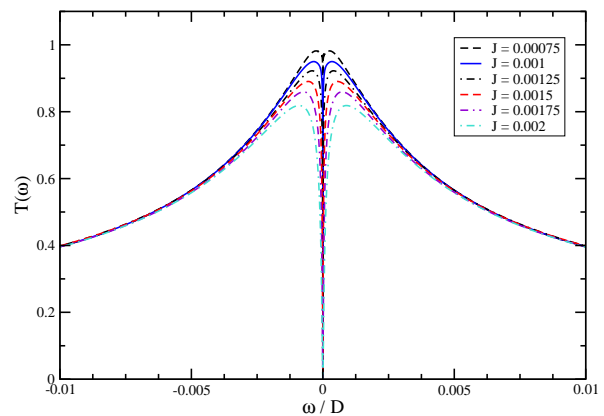}
\end{center}
\caption{Transmission coefficient through dot 
$1$ at zero temperature for different RKKY couplings, $J$. 
The energy unit is the half bandwidth, $D=1$. 
The Anderson model's parameters were   $U=1$, 
$\protect\epsilon_{d 1}=\protect\epsilon_{d 2}=-0.5$ , 
$\Gamma_{1L}=\Gamma_{1 R}=0.1$, resulting in a Kondo temperature of 
$T_K\approx 0.0055$. We used a discretization parameter 
$\Lambda=2$. 
}
\label{TwJ}
\end{figure}

{\em Comparison with NRG.} The transport properties  of the
side-coupled quantum dot system can also be studied numerically by 
Numerical Renormalization Group (NRG) methods. 
The transmission coefficient $T(\omega )$ through dot $1$
 obtained from NRG at different RKKY couplings $J$ is plotted 
in Fig.\ref{TwJ}. In all figures we compensated for a 6\% loss of the spectral
weight. The large resonance is a manifestation of the Kondo effect
displayed by the first dot, and the sharp dips  in the transmission at 
$\omega\approx 0$ are due to the formation of the singlet state below the energy 
$T^\ast$.  

\begin{figure}[h]
\begin{center}
\includegraphics[width=0.9\columnwidth,clip]{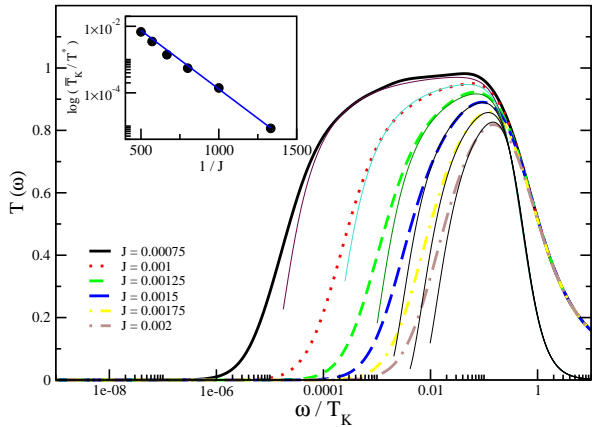}
\end{center}
\vspace{-0.1cm}
\caption{Fit of the numerically obtained transmission coefficient 
$T(\protect\omega)$ by the perturbative expression, \eqref{eq:G_d}
for various values of $\tilde J$.  For all fits we used $\tilde T_K = 0.5 T_K
= 0.003$.  The inset shows the scale $T^\ast$
extracted from the fit, as a function  of $1/J$.}
\label{fit}
\end{figure}

The numerically obtained 
transmission coefficients are compared to the analytical formula \eqref{eq:G_d}
in Fig.~\ref{fit}. The perturbative expression agrees well with the numerical
results over a wide range of energy scales between $T^\ast $ and $T_K$.
It is interesting to observe the deviations {\em above } $T_K$, 
where  the simple Lorentzian approximation we made 
for $G_d^{(0)}$ fails to account for the fact that
the resonance on dot 1 is also a Kondo resonance. This Lorentzian approximation
thus completely neglects the large logarithmic tails 
for $\omega>T_K$.  
In the inset we also show the scale $T^\ast$ as extracted 
from our fits as a function of $1/ J$. The extracted scales  compare 
very well with the  analytical expression, Eq.~\eqref{eq:Tstar}, 
indicated by the solid line.

\begin{figure}[pt]
\begin{center}
\includegraphics[width=0.95 \columnwidth,clip]{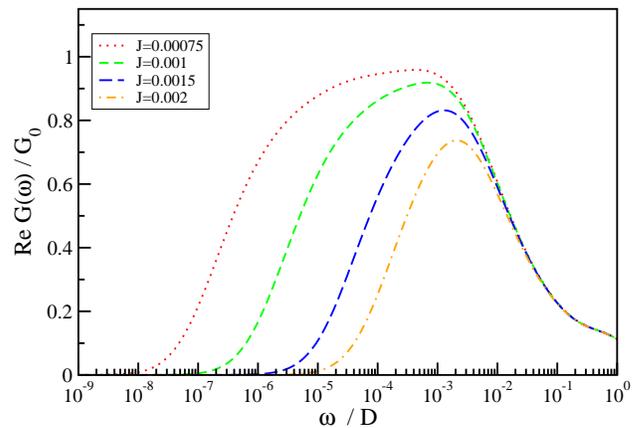}
\end{center}
\vspace{-0.1cm}
\caption{
AC conductance normalized by $G_0$ 
for the same  parameters as in Fig.~\protect\ref{TwJ}. 
}
\label{GwJ}
\end{figure}

The two-stage Kondo effect  and the K-T transition can also 
be observed in the AC conductance of the two-dot device or 
the noise spectrum. Such measurements have been indeed 
done in recent experiments of high-frequency current
fluctuations\cite{deblock},   though the linear AC conductance measurements in 
the relevant frequency regime are very difficult due to background currents 
from the capacitors.\cite{nordlander} According to Eq.~\eqref{eq:ReG}, the 
real part of the conductance through the device can be computed
from the transmission coefficient through a simple integration. The resulting 
curves are displayed in Fig.~\ref{GwJ}.

\begin{figure}[bp]
\begin{center}
\includegraphics[width=0.95\columnwidth,clip]{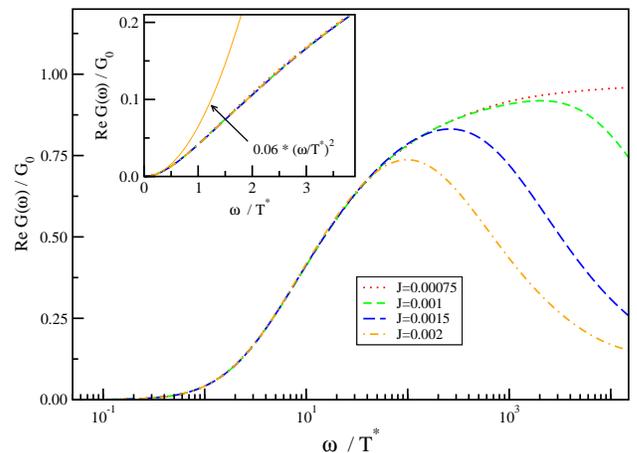}
\end{center}
\vspace{-0.1cm}
\caption{
$G'(\omega)$ as a function of $\omega/T^\ast$ for various values 
of $J$. Below the Kondo scale all curves collapse to a 
universal curve. Inset: small frequency part of this 
universal curve on a linear scale. We also show the quadratic behavior
characteristic of electron-hole symmetry, determined from a 
log-log fit of the curves.}
\label{fig:universal}
\end{figure}

It is interesting to remark that the low energy cross-over of the transmission
coefficient and the conductance is described by {\em universal cross-over functions} 
for $\omega,T^\ast \ll T_K$. The conductance, e.g., is approximately given by 
\bea
G'(\omega) &\approx&  G_0 \; g(\omega/T^\ast)\;
\eea
in this regime.
Here the scaling function $g$ depends somewhat on electron-hole symmetry breaking, 
but in case of electron-hole symmetry it is  completely universal. Then for very small
frequencies  it  scales to zero as $g(\omega) \approx 0.06\; (\omega/T^\ast)^2 $, 
while  at high
energies it approaches   1 logarithmically, $g(\omega/T^\ast) \approx 1 -
\alpha/\log^2(\omega/T^\ast)$. This universal cross-over function can be
extracted from the NRG results, and is displayed in Fig.~\ref{fig:universal}. 
The transmission coefficient $T_\sigma(\omega)$ displays similar universal
scaling properties.

\begin{figure}[h]
\begin{center}
\includegraphics[width=0.9\columnwidth,clip]{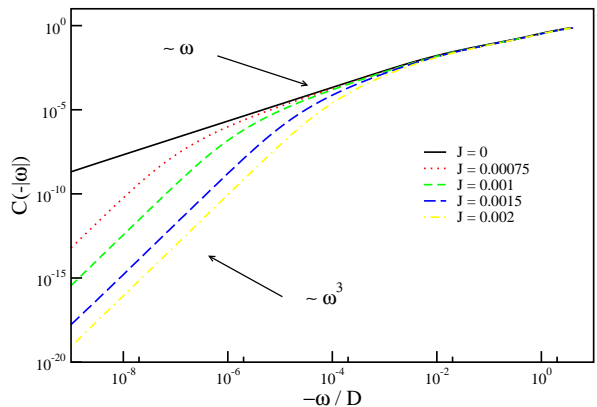}
\end{center}
\vspace{-0.1cm}
\caption{
Equilibrium current noise $C(\omega )$ in case of electron-hole symmetry 
for $\omega <0$. 
$C(\omega )$  exhibits linear dependence 
on  $\omega $ for $T^\ast< |\omega| <T_K$, that crosses over to 
a cubic behavior below $T^\ast$.  Note that $C(\omega )=0$ for 
$\protect\omega >0$. The parameters are the same as in Fig.\protect\ref{TwJ}.
}
\label{CwJ}
\end{figure}

Finally, let us discuss  the current noise at zero temperature, plotted in Fig.\ref{CwJ}. 
At $T=0$ temperature $C(\omega)$ has only weight for $\omega<0$.
For  $T^\ast< -\omega<T_K $ we have $C(\omega )\propto |\omega |$ 
corresponding to the Fermi-liquid property of a perfectly transmitting 
quantum dot.\cite{sindel} However, for $-\omega< T^\ast$ 
this behavior crosses over to a power law scaling, where in case of
electron-hole symmetry  one has $C(\omega )\propto |\omega |^3/(T^\ast)^2$. 
This behavior is somewhat modified once electron-hole symmetry is
broken. Then the conductance remains finite even in the $\omega\to 0 $ limit, 
and correspondingly, another cross-over may take place from the 
 $C(\omega )\propto |\omega |^3 $ regime to a 
linear regime, $C(\omega )\propto |\omega |$ at some energy 
$T^{\ast\ast}\ll T^\ast$.

{\em Conclusions.} We provided a complete analytical and numerical analysis of
the AC  transport properties and the two-stage Kondo screening in 
side-coupled quantum dots.  Our analytical results were based on
renormalized  perturbation and scaling theory, and they  agree well with 
the NRG result  over a wide frequency range. We also determined  the linear 
AC conductance  and the equilibrium current noise which can both be measured 
experimentally and reflect the two-stage Kondo effect and the
K-T transition. We also computed the universal cross-over functions
that describe the emergence of the triplet state at the energy 
$T^\ast$. 
 
{\em Acknowledgments.} This work has
been supported by the DFG-Center for Functional Nanostructures and by the 
Virtual Institute for Research on Quantum Phase Transitions at the University 
of Karlsruhe. C.H.C acknowledges the support from the NSC and the MOE 
ATU Program of Taiwan, R.O.C. G.Z. has been supported by the Humboldt
Foundation and by Hungarian grants, OTKA Nrs. NF061726, and T046303. 
He also acknowledges the hospitality of the C.A.S., Norway, where part of this
research has been done.


\begin{references}
\bibitem{footnote}  This cut-off is typically provided
by the level spacing $\delta \epsilon$ of the quantum dots.
\end{references}
\end{document}